\documentclass{IEEEcsmag}

\usepackage[colorlinks,urlcolor=blue,linkcolor=blue,citecolor=blue]{hyperref}
\expandafter\def\expandafter\UrlBreaks\expandafter{\UrlBreaks\do\/\do\*\do\-\do\~\do\'\do\"\do\-}
\usepackage{upmath,color}
\usepackage{longtable}
\usepackage{booktabs}
\let\labelindent\relax
\usepackage{enumitem}
\usepackage{multirow}
\usepackage{tabularx}
\usepackage{float} 

\jvol{XX}
\jnum{XX}
\paper{8}
\jmonth{May}
\jname{METACOG-25}
\jtitle{}
\pubyear{2025}

\begin{document}

\sptitle{Second Workshop on Metacognitive Prediction of AI Behavior}

\title{Exploring Cognitive Attributes in Financial Decision-Making}

\author{Mallika Mainali}
\affil{Drexel University, Philadelphia, PA, 19104, USA}

\author{Rosina O. Weber}
\affil{Drexel University, Philadelphia, PA, 19104, USA}

\markboth{}{}

\begin{abstract}\looseness-1Cognitive attributes are fundamental to metacognition, shaping how individuals process information, evaluate choices, and make decisions. To develop metacognitive artificial intelligence (AI) models that reflect human reasoning, it is essential to account for the attributes that influence reasoning patterns and decision-maker behavior, often leading to different or even conflicting choices. This makes it crucial to incorporate cognitive attributes in designing AI models that align with human decision-making processes, especially in high-stakes domains such as finance, where decisions have significant real-world consequences. However, existing AI alignment research has primarily focused on value alignment, often overlooking the role of individual cognitive attributes that distinguish decision-makers. To address this issue, this paper (1) analyzes the literature on cognitive attributes, (2) establishes five criteria for defining them, and (3) categorizes 19 domain-specific cognitive attributes relevant to financial decision-making. These three components provide a strong basis for developing AI systems that accurately reflect and align with human decision-making processes in financial contexts.
\end{abstract} 
\maketitle
\chapteri{A}s Artificial Intelligence (AI) decision-making models and recommender systems(RSs) are increasingly deployed in high-stake applications, various metacognitive AI frameworks have been proposed to increase the explainability of these models.\textsuperscript{\cite{wei2024} \cite{tankelevitch2023} \cite{zhong2024}} However, the issue of ensuring that these AI models align with human behavior and values that shape our decision-making process is still one of the most significant scientific issues in AI research.\textsuperscript{\cite{christian2021}} To develop metacognitive AI models capable of capturing uncertainties in decision-making, it is crucial to account for the individual attributes that shape human reasoning.\textsuperscript{\cite{SMOLINSKI2024} \cite{Waytz2010}} By systematically integrating these attributes, AI models can better represent complex cognitive patterns,\textsuperscript{\cite{Bendell2021}} enhancing their predictive and adaptive capabilities.\textsuperscript{\cite{angulo2020} \cite{Beheshti2020}} 

A fundamental limitation of current AI research is its lack of integration with human cognitive attributes.\textsuperscript{\cite{Theocharous2019}} Decision-making in humans is not entirely rational\textsuperscript{\cite{Stanovich2000}} and is influenced by complex cognitive processes, including affect,\textsuperscript{\cite{Loewenstein2003}} heuristics,\textsuperscript{\cite{Tversky1974}} and bounded rationality,\textsuperscript{\cite{Simon1990}} all of which vary between individuals. 
Some work has been carried out in aligning AI models with human attributes, such as the Multiobjective Maximum Expected Utility paradigm that combines vector utilities and non-linear action selection to address limitations in aligning AI effectively,\textsuperscript{\cite{Vamplew2018}} and Trustworthy Algorithmic Delegate (TAD) which aligns decision-making with target decision-makers by using case-based reasoning, Monte Carlo simulation, Bayesian diagnosis, and naturalistic decision-making.\textsuperscript{\cite{Molineaux2024}} 
\setlength{\parskip}{0pt} 

In consequential domains such as behavioral finance, where cognitive attributes are essential for aligned reasoning,\textsuperscript{\cite{Takayanagi2024}} there is currently no standard methodology for gathering domain-specific cognitive attributes. This study addresses this gap by exploring the influence of cognitive attributes in decision-making. By doing so, we seek to inform the design of aligned AI models for real-world applications. To limit the scope of our study, we focus on finance, a consequential domain with universal applicability, where integrating cognitive attributes can help develop objective and effective AI systems.
Thus, our study has the following objectives:
 
\begin{itemize}
\item[{\ieeeguilsinglright}] Explore the origins of cognitive attributes by examining their foundations in cognition and decision-making literature.
	
\item[{\ieeeguilsinglright}] Propose a set of criteria to define cognitive attributes that can be transferable across domains.

\item[{\ieeeguilsinglright}] Identify key cognitive attributes relevant to financial decision-making by examining their influence on individuals' financial choices.
\end{itemize}\vspace*{3pt}
\setlength{\parskip}{0pt} 
\vspace{-2mm}
Our key contributions include: (1) an analysis of the origins of cognitive attributes, (2) criteria for defining cognitive attributes, and (3) a literature review on key cognitive attributes relevant to financial decision-making and their impact on financial decisions.


\vspace{-5mm}
\section{Origins Of Cognitive Attributes}
The study of cognitive attributes has its roots in philosophy,\textsuperscript{\cite{Thagard2023}} but its comprehension has evolved through research in psychology and neuroscience. Wundt’s formalization of introspection in the study of the brain was one of the first scientific explorations of cognition,\textsuperscript{\cite{Titchener1920}} followed by the cognitive revolution, which led to the development of cognition information processing models.\textsuperscript{\cite{Atkinson1968} \cite{Baddeley1974}} This revolution shifted attention from the study of behaviorism to understanding internal mental processes and creating representational models of different components of processing, such as attention, memory, and decision-making.\textsuperscript{\cite{Neisser1967}}

In this era of cognitive revolution, Neisser introduced the concept of \textit{cognitive processes}, which are the mental mechanisms that influence how we perceive, reason, and make decisions.\textsuperscript{\cite{Neisser1967}} He illustrated that cognitive processes are constructive and contextual functions rather than passive reactions. Similarly, in contrast to the fixed representation of cognitive attributes demonstrated in earlier cognitive science research, Piaget introduced the theory of cognitive development to explain how children's reasoning abilities evolve through four stages: sensorimotor intelligence, preoperational thinking, concrete operational thinking, and formal operational thinking.\textsuperscript{\cite{Piaget1971}} He highlighted the dynamic role of cognitive attributes by defining \textit{constructivism} and illustrating that children actively build their knowledge by engaging with the world around them and refining their cognitive structures as they grow.\textsuperscript{\cite{Piaget1964}} Tversky and Kahneman further explored how cognitive attributes influence decision-making by introducing cognitive heuristics as mental shortcuts that simplify decision-making.\textsuperscript{\cite{Tversky1974}} They introduced three heuristics: \textit{representativeness, availability}, and \textit{adjustment \& anchoring}, and concluded that these heuristics often lead to cognitive biases that influence human judgment. Similarly, Simon’s theory of bounded rationality further explained how cognitive and environmental limitations restrict individuals from making rational decisions.\textsuperscript{\cite{{Simon1955}}} He demonstrated that people tend to \textit{satisfice} rather than optimize, meaning they often choose an option that meets acceptable criteria instead of exhaustively searching for the best possible option.

Due to their ability to vary based on environmental contexts, cognitive attributes were regarded as semi-stable and dynamic patterns influenced by context,\textsuperscript{\cite{Bechara1997}} as described in models such as the \textit{Cognitive-Affective Personality System} (CAPS).\textsuperscript{\cite{Mischel1995}} The CAPS model challenged the idea of fixed personality traits and emphasized the role of situational contexts in shaping behavior by focusing on how cognitive and affective processes, such as goals, expectations, and emotions, interact with environmental contexts to produce consistent behavioral patterns. 

In behavioral economics, \textit{Prospect Theory} explained how cognitive attributes such as loss aversion and risk preferences affect decision-making.\textsuperscript{\cite{Kahneman1979}} Instead of assuming that individuals make decisions based on rational calculations of expected utility, this theory demonstrated how human decisions are influenced by cognitive biases and psychological factors.\textsuperscript{\cite{Kahneman1979}} Similarly, Thaler's \textit{theory of mental accounting} described how individuals compartmentalize money based on attributes such as endowment effect, loss aversion, and self-control.\textsuperscript{\cite{Thaler1980}} His work explained the human tendency to deviate from rational decision-making and to separate finances into different categories, which in turn affected their spending behavior. 

Based on the work of Neisser, Piaget, Tversky, Kahneman, and Simon, we observe that cognitive attributes are now recognized as dynamic and context-dependent. The term \textit{cognitive attributes} has evolved into an umbrella term that incorporates a range of individual differences (\textit{e.g.,} heuristics, biases, decision-making styles) that influence decision-making.
\vspace{-8mm}
\section{Criteria For Cognitive Attributes}
As discussed in the previous section, cognitive attributes influence perception, reasoning, and decision-making, shaping how individuals navigate uncertainty. To define cognitive attributes, we propose the following criteria:
\begin{enumerate}
\setlength{\topsep}{0pt} 
\item \textbf{Rooted in Cognitive Psychology:} 
A cognitive attribute must be recognized in psychological research as a trait that influences how individuals acquire information, design alternative strategies, and make decisions.
\item \textbf{Contextual and Semi-Stable:} 
Cognitive attributes are contextual and semi-stable. They tend to remain consistent over time\textsuperscript{\cite{Frey2017}} but can fluctuate with context, experience, or cognitive load.\textsuperscript{\cite{Stanovich2000}} For example, an individual’s risk tolerance may shift due to significant life events, but it tends to remain relatively stable overall.\textsuperscript{\cite{VandeVenter2012}} While personality traits can also change, they are generally less sensitive to immediate context and short-term fluctuations than cognitive attributes.\textsuperscript{\cite{Roberts2006}}

\item \textbf{Based on metal processes:} 
Cognitive attributes stem from thought processes and stable cognitive patterns that guide how individuals evaluate options and make decisions. While emotions like fear or stress can influence these attributes, they are not cognitive attributes themselves as they are transient, reactive, and tied to affective processes rather than deliberative cognition.\textsuperscript{\cite{Slovic2004}} 
This makes emotions more unstable than cognitive attributes.

\item \textbf{Influences Decision-Making and Behavior:} 
Cognitive attributes affect how individuals interpret situations, assess outcomes, and make choices. By influencing how past experiences are applied to current situations, they shape both daily decisions and long-term behaviors. So, traits like age, gender, physical characteristics, emotions, and external factors such as weather or location are not cognitive attributes since they do not directly guide cognitive processes in evaluating options or applying past knowledge.

\item \textbf{Consistently Measurable:} 
Cognitive attributes should be observable and quantifiable across contexts, allowing for reliable assessment and prediction of decision-making patterns. For example, risk tolerance can be measured through standardized questionnaires, while temporary emotional states cannot be consistently quantified.
\end{enumerate}
\vspace{-5mm}
\section{Cognitive Attributes In Finance}
Financial decision-making (FDM) is inherently influenced by an individual’s risk perception, tolerance for uncertainty, and the complexity of the environment in which decisions are made.\textsuperscript{\cite{Frydman2016}}
\begin{table}
\vspace*{2pt}
\caption{Cognitive Attributes Influencing Financial Decisions}
\label{tab:financial cognitive attributes}
\tablefont

\begin{tabular*}{17.5pc}{@{\extracolsep{\fill}}p{87.5pt}<{\raggedright}p{87.5pt}<{\raggedright}@{}}

\toprule
Category & Cognitive Attributes \\
\colrule
Risk Perception & 
Risk tolerance \\
& Risk aversion \\[3pt]

Decision Processing & Numeracy \\
Style & Temporal discounting \\ 
& Domain knowledge \\
& Cognitive reflection \\
& Conscientiousness \\
& Mental accounting \\
& Perceived ownership \\
& Locus of control \\
& Self-control \\[3pt]

Emotional and Social & Confidence \\
Influences & Optimism \\
& Regulatory focus \\
& Herding behavior \\[3pt]

Cognitive Biases &  Default bias \\
& Selection bias \\
& Loss aversion \\
& Conjunction fallacy \\
\botrule
\end{tabular*}
\vspace*{-9mm}
\end{table}

\begin{figure*}
\centerline{\includegraphics[width=26pc]{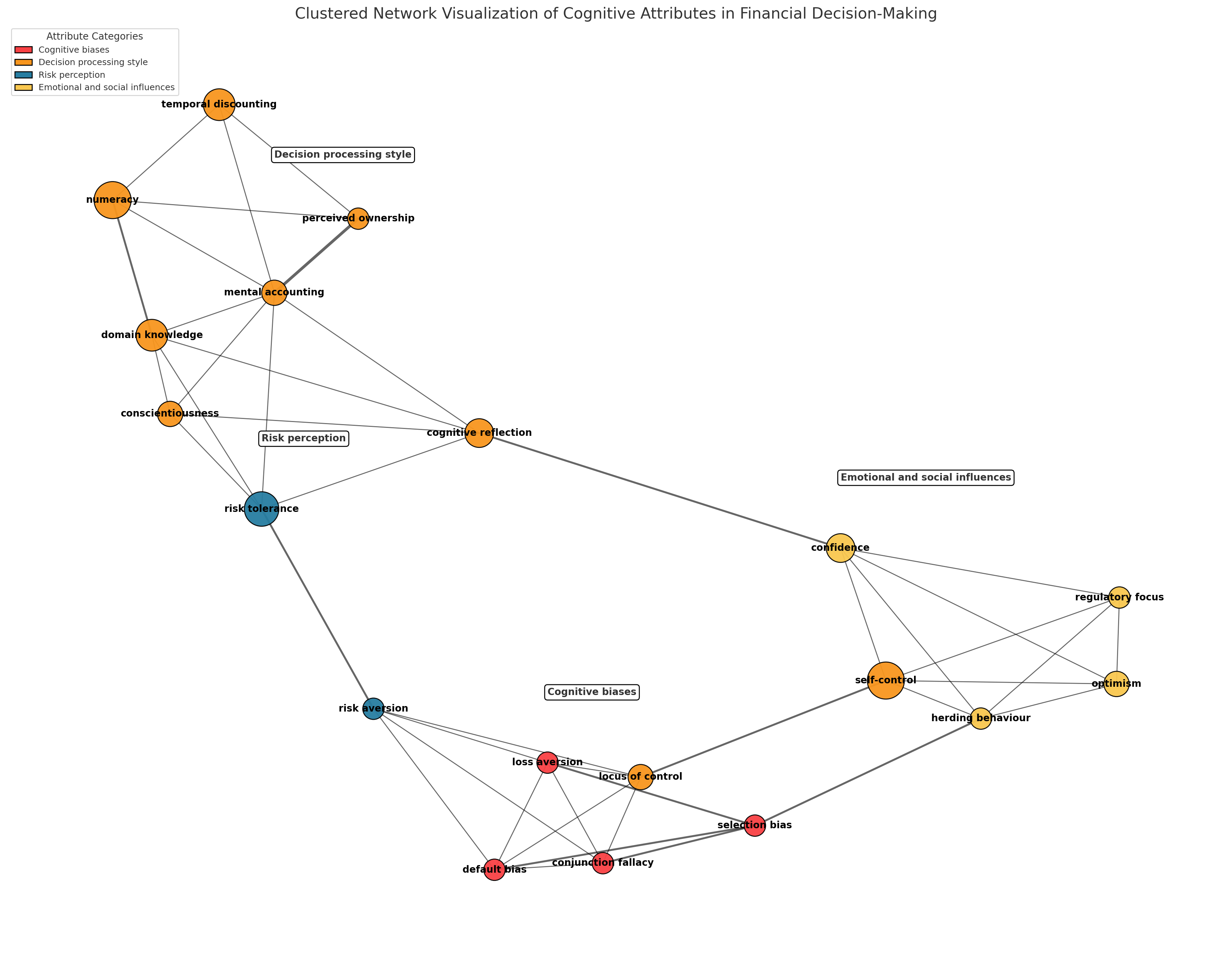}}
\vspace*{-23pt}
\caption{Clustered network visualization of cognitive attributes influencing financial decision-making}
\label{fig:cognitiveattributes}
\vspace*{-15pt}
\end{figure*}
In our evaluation of existing literature, we focused on cognitive attributes that shape financial choices. To identify them, we conducted a structured literature review using the Scopus database. We employed the following search query:

\textit{TITLE-ABS-KEY(("cognitive attributes" OR "personal attributes" OR "cognitive traits" OR "individual attributes" OR "individual traits" OR "individual differences") AND ("financial decision-making"))}

This search returned 40 articles, all of which explicitly addressed individual attributes in the context of FDM. After reviewing the abstracts for relevance, we included all 40 articles for further analysis and extracted the cognitive attributes referenced in them, which resulted in the extraction of 71 attributes. We further reduced them to 46 distinct attributes after removing duplicates. Each of these 46 attributes was manually assessed against the five predefined criteria. In this process, we excluded non-psychological attributes such as age, gender, education level, and income. We also standardized terminology across articles. For instance, we replaced \textit{cognitive ability} with \textit{cognitive reflection} since they referred to the same concept in our context, and grouped related terms like \textit{domain education} and \textit{domain experience} under \textit{domain knowledge}. This process resulted in 35 cognitive attributes. After removing duplicates resulting from synonyms, we had a final set of 19 distinct cognitive attributes that satisfied all five criteria. These 19 attributes form the basis for our subsequent analysis. We found that numeracy, self-control, risk tolerance, temporal discounting, and domain knowledge were the most prominent cognitive attributes across the literature. The complete process and details of attribute selection are provided in the appendix.
Figure \ref{fig:cognitiveattributes} presents a clustered network visualization of the final cognitive attributes, where larger nodes indicate more frequently referenced attributes.

We categorized these attributes into four areas: \textit{Risk Perception, Decision Processing Style, Emotional and Social Influences}, and \textit{Cognitive Biases}. {\it Risk perception} involves how individuals assess financial risks, while {\it decision processing style} explores how individuals process information and apply relevant knowledge. {\it Emotional and social influences}, along with {\it cognitive biases}, refer to the affective, social, and systematic errors that shape decision-making. A complete list of identified attributes is provided in Table \ref{tab:financial cognitive attributes}.
\vspace{-8mm}
\section{Influence On Financial Decisions}
In this section, we examine how each of the 19 cognitive attributes identified in our literature review influences financial decision-making. 

We found that higher {\it risk tolerance} was associated with more aggressive investment strategies,\textsuperscript{\cite{Grable2000}} while {\it risk aversion} correlated with safer investments.\textsuperscript{\cite{Bernoulli1954}} Similarly, higher {\it numeracy} was linked to higher net worth,\textsuperscript{\cite{Yilan2021}} and stronger {\it domain knowledge} reduced reliance on heuristics and biases.\textsuperscript{\cite{James2012} \cite{Guo2022}} Likewise, individuals with higher {\it cognitive reflection} were more likely to make rational financial decisions, as they often replace intuitive responses with deliberate reasoning.\textsuperscript{\cite{Frederick2005}}
Individuals with higher {\it temporal discounting} favored short-term gains over long-term benefits, often leading to impulsive or suboptimal decisions.\textsuperscript{\cite{Critchfield2001}} Similarly, low {\it self-control} contributed to impulsive purchases and poor financial planning,\textsuperscript{\cite{Otto2010}} whereas {\it conscientiousness} promoted disciplined and goal-oriented behavior.\textsuperscript{\cite{Yilan2021}}

{\it Mental accounting} led to inconsistent financial decisions as people categorize money differently based on its source or intended use,\textsuperscript{\cite{Mahapatra2022}} while {\it perceived ownership} created a psychological attachment to assets, causing individuals to hold on to underperforming investments due to the endowment effect.\textsuperscript{\cite{Kahneman1991}} 

We also found that individuals with an internal {\it locus of control} took more proactive financial actions.\textsuperscript{\cite{Salamanca2020}} In terms of {\it regulatory focus}, individuals with promotion focus were found to seek more gains and take more risks compared to prevention focus.\textsuperscript{\cite{Kumar2016}} Additionally, while moderate {\it confidence} positively influenced decision-making, overconfidence resulted in excessive trading and risk-taking.\textsuperscript{\cite{Bouteska2023}} Likewise, {\it optimism} increased risk-taking. 
{\it Herding behavior} led individuals to follow others’ financial actions, ignoring their analysis,\textsuperscript{\cite{BikhchandaniSharma2000}} while {\it default bias} kept them anchored to pre-set options (\textit{e.g.}, default retirement plans). {\it Selection bias} skewed decisions by relying on non-representative information, leading to flawed judgments.\textsuperscript{\cite{Koehler2009}} {\it Loss aversion} made individuals fear losses more than they valued gains, causing cautious behavior.\textsuperscript{\cite{Dolder2024}} Lastly, {\it conjunction fallacy} caused reasoning errors where individuals overestimated the probability of specific scenarios. 
\vspace{-8mm}
\section{Conclusion}
In this study, we explored economic {\it (e.g.,} Prospect Theory) and psychological {\it (e.g.,} bounded rationality) theories that contribute to understanding decision-making processes. Building on these theoretical principles, we established five criteria for defining and identifying cognitive attributes and identified 19 cognitive attributes relevant to financial decision-making. Our goal was to emphasize the importance of incorporating cognitive attributes in developing AI models, enabling them to predict and recommend financial decisions that align with human needs and preferences. This approach provides a basis for designing AI systems that can leverage cognitive attributes to personalize recommendations or serve as an integral component in the reasoning processes of metacognitive AI approaches. Future work could explore their application in real-world contexts to improve AI alignment.

\vspace{-7pt}
\section{Acknowledgments}
This research was conducted as part of the In the Moment (ITM) project, supported by the Defense Advanced
Research Projects Agency (DARPA) under contract number
HR001122S0031.

\vspace{-7pt}
\def\refname{References}

\vspace*{-8pt} 
\newpage
\section{Appendix}
A review of 40 articles selected from Scopus resulted in the extraction of 71 attributes relevant to financial decision-making. From this set, we identified 46 distinct attributes by grouping conceptually similar attributes and eliminating duplicates. The list of these 46 attributes, before any refinements, is presented in the bar chart in Figure \ref{fig:71attributes}. These attributes were further refined based on the process described below:
\begin{itemize}
    \item[{\ieeeguilsinglright}] \textbf{Refined terminology for consistency:}  
    \begin{itemize}
        \item[{\ieeeguilsinglright}] \textit{Anticipation} was replaced with \textit{optimism} based on literature supporting their correlation.  
        \item[{\ieeeguilsinglright}] \textit{Cognitive ability} was replaced with \textit{cognitive reflection} as the referenced article used them interchangeably.  
        \item[{\ieeeguilsinglright}] \textit{Delayed gratification} was replaced with \textit{self-control}, supported by existing research on their conceptual alignment.  
        \item[{\ieeeguilsinglright}] \textit{Domain education, domain experience, and domain literacy} were grouped under \textit{domain knowledge} to consolidate similar concepts.  
        \item[{\ieeeguilsinglright}] \textit{Financial responsibility and Persistence} were replaced with \textit{conscientiousness}, as their conceptual components were synonymous in our context.  
        \item[{\ieeeguilsinglright}] {\it Impulsivity} was removed as it is inversely related to {\it self-control} and closely associated with {\it temporal discounting}.
    \end{itemize}
    \item[{\ieeeguilsinglright}] \textbf{Removed attributes that did not meet predefined criteria:}  
    \begin{itemize}
        \item[{\ieeeguilsinglright}] \textit{Age, gender, education level, sociodemographic characteristics, household income, personal income, and subjective life expectancy} were eliminated as they are not psychological attributes and are not rooted in cognitive psychology.
        \item[{\ieeeguilsinglright}] \textit{Personality trait} was removed as it is a broad construct and generally more rigid than cognitive attributes.  
        \item[{\ieeeguilsinglright}] \textit{Overreaction \& underreaction, and stress} were excluded as they are highly subjective, lack the stability needed to constitute a cognitive attribute, and are more indicative of emotional regulation than cognitive processes.  
        \item[{\ieeeguilsinglright}] \textit{Investment sentiment} and constructs reflecting the collective mood or psychological state of groups were removed.  
        \item[{\ieeeguilsinglright}] \textit{Spending habits} and \textit{attitude towards savings} were excluded, as they are not rooted in cognitive psychology but are instead shaped by personal and social factors.  
    \end{itemize}
\end{itemize}

After completing this refinement process, we reviewed the final set of attributes to ensure that they satisfied the five predefined criteria. Out of the initial 46 attributes, we identified 35 cognitive attributes that satisfied our criteria. After removing duplicates, we narrowed this down to 19 distinct attributes that met all five criteria. These were used in our analysis to examine their influence on financial decision-making, which revealed a direct impact on financial choices and highlighted their relevance in designing recommendation, prediction, and reasoning models.

\end{document}